# Generation and validation of custom multiplication IP blocks from the web

Minas Dasygenis (mdasyg@ieee.org)
Department of Informatics and Telecommunications Engineering,
University of Western Macedonia, Kozani, 50100, Greece

*Abstract*—Every CPU carries one or more arithmetical and logical units. One popular operation that is performed by these units is multiplication. Automatic generation of custom VHDL models for performing this operation, allows the designer to achieve a time efficient design space exploration. Although these units are heavily utilized in modern digital circuits and DSP, there is no tool, accessible from the web, to generate the HDL description of such designs for arbitrary and different input bitwidths. In this paper, we present our web accessible tool to construct completely custom optimized multiplication units together with random generated test vectors for their verification. Our novel tool is one of the firsts web based EDA tools to automate the design of such units and simultaneously provide custom testbenches to verify their correctness. Our synthesized circuits on Xilinx Virtex 6 FPGA, operate up to 589 Mhz.

## I. INTRODUCTION

The design automation and test processes (DAT) play a crucial role in contemporary multi-billion transistor era of heterogenous computing. One of the aspects of DAT is the fast parametrized generation of bit accurate models and their test vectors, in a hardware description language (HDL). This enables the designers to perform a rapid design space exploration and select the best custom implementation for their embedded systems. Especially, the HDL generators for constructing circuits that are required in almost every digital system have an increased importance. This is even more true for designing custom heterogenous architectures.

Multiplication is an important elementary operation in the image processing domain within digital systems [20]. Every modern microprocessor has this operation within its instruction set. Especially, this operation is valuable in digital signal processing (DSP) applications, like Image processing or Discrete Fourier Transform [17], where it is a fundamental operation in these algorithms. For this reason, performance of any DSP processor is defined with delays in multiply and accumulate (MAC) units.

In the literature, there are a lot of ways to perform a multiplication [3], [5], [15], [21] and many other. One very effective technique to perform multiplication is the creation of the partial sums using a logic gates network and the addition of these bits with multiple vectors in a carry save manner (CSA). Even though multiplication is a very common operation, nobody has created a web accessible tool to create parametrized multiplication units in a hardware description language, using CSA. Of course major EDA vendors, like Xilinx or Altera provide a functionality to create parametrized IP multipliers block, but these IP descriptions are created in an encrypted binary form that is device locked to their boards, and sometimes cannot be freely used without a license.

We noticed this shortcoming and decided to create a tool that will be able to create generic two vector multipliers, with parametrized input bitwidths with or without pipeline to be used for custom heterogenous architectures, like image processing accelerators. To the best of our knowledge, our tool is the first web based EDA tool to provide custom IP multiplier modules for different input bitwidths and provide random testbenches to verify them. Thus, our major contributions of our work are: ($a$) we present a public web accessible tool that can create very fast syntactically correct register-transfer-level VHDL description of a multiplier according to user inputs, ($b$) we demonstrate the usefulness of using a high level language (Python) in the rapid design space exploration of embedded systems, ($c$) we present a web EDA tool in order to motivate other designers or vendors to provide such functionality from the web, and move the electronic design automation from the desktop to the web.

The rest of the paper is organized as follows. The next section (Section II) discusses some related published work on multipliers and automatic generation of HDL codes. Section III presents our multiplier tool, while Section IV presents metrics and experimental results of the tool and the generated codes. Finally, we give the concluding remarks in Section V.

## II. RELATED WORK

Multiplication is a common operation, much more complicated than addition. For this reason, the design of high speed multiplier has always played a significant role in the EDA landscape for over 4 decades. Even from 40 years ago, authors [4] have tried to optimize this operation and provide ways to decrease the cycles required. In the research databases, we can find a large number of publications on this subject that use many schemes, like tree multipliers or array multipliers. ([3], [5], [15], [21], are a few of them). The general opinion is that tree multipliers are faster, but harder to design, while on the other hand, multipliers that use arrays of bits have a better layout. One of the array multiplier techniques, is the one that uses carry save adders. For this reason, we decided to create a generator for creating multiplication IP blocks using carry save adders.

After deciding the technique to implement our module, we noticed that in the web there was (and still is) a lack of online EDA applications. Even though, during the previous years, many applications and work flows have been transferred to the cloud, there is a significant gap for the EDA applications. Researchers have noted this from 12 years ago [18] and had pinpointed the benefits of web based tools. Currently, from the major EDA companies, Synopsis provides the web VCS tool for verification. On the literature anybody can locate plenty of EDA tools. For example, Schneider et al [19] described their Java tool that accepts as input a special form of statements and generates VHDL for the arithmetic circuits. The tool is not available to the public, and from the published results we can see that the generated multiplication units have very low operation frequency (only 101Mhz), due to the absence of the pipelining and because the circuits are restricted to on-line arithmetic (all operations are performed digit serially). Another EDA tool was published 20 years ago from the Mentor Graphics engineers Kumar et al [14], which accepts a special input stream of arithmetic statements and creates a netlist specific for the tool Mentor Graphics Autologic. The user cannot fine tune this, like using pipeline or defining the input bitwidths, and again the tool is not available. EDA tools have also being created to optimize specific implementation technologies. Pihl et al [16], Wen-Jong [8] and Abke et al [1] created tools to optimize arithmetic circuits either for FPGA or for VLSI ASIC design flows. They used Wallace Trees and Booth Encoding [16], clever use of the CLB multiplexors [1], and utilizing the resources of every FPGA macrocell [8]. Tools for High Level Synthesis have been also published by Zhi et al [10] in their ROCCC suite, in which a C description is synthesized into HDL code, and by Hannig et al [12] in their PARO compiler, in which an elaborate description in PARO is implemented in systolic architectures.

A similar work with ours has been recently presented by Florent de Dinechin et al [7], in their tool Flopoco, which can generate VHDL code for arithmetic cores. Compared to our work, their work is not online and must be downloaded and compiled, together with all the library dependencies, something that is time consuming, requires root privileges and knowledge on the Linux administration about configuring products, resolving library conflicts, satisfying package dependencies, compiling and linking. Also, their integer multiplication units utilize binary compressors specific for the DSP blocks of an FPGA, and they are not architectural neutral, like our implementation. We provide comparison estimations of ours and Flopoco's generated architectures on Section IV.

Another interesting and similar tool to our work was published by Bakalis et al [2]. The authors provide a tool, via a web interface, which delivers HDL code for some arithmetic modules (including multiplication). The delivered circuits are not pipelined, the operands on the multiplication must be equally sized, due to the two-bit recoding and the operand width is limited to 256 bits. The limitation of creating multipliers of having always the same operand bitwidth is very restricting and does not allow fine tuning the module for the exact design requirements.

Finally, another web tool has been presented by Voronenko et al [22], but it is only limited in designing custom HDL for Multiple Constant Multiplication problems, and not for generic vector multiplication.

With the exception of [2], which yields some limitations, due to the implemented multiplier design, no other has ever presented a web based tool to create generic multiplier blocks from the Internet, in order to help designers perform a quicker design space exploration. For this reason we provide such a tool, which implements one of the most popular multiplication algorithms (CSA multiplication), lifting the drawbacks noted on the previous tools. Also, our tool is the only tool to provide random test vectors to the end-user to verify its correct operation.

III. THE WEB MULTIPLIER COMPILER

Our tool, which has been installed on a public web server[1], utilizes a number of technologies (PHP, Python, JSON), in order to deliver a syntactically correct and synthesizable VHDL description. Our tool is partitioned in two different departments, according to their function: the front end and the back end. These modules exchange information using the Javascript object notation (JSON) format [9].

Concerning the high level language selection of Python, we can justify our decision by the following. Traditionally, the EDA landscape was dominated by legacy and proprietary languages (LISP, PERL, TCL, dc-shell, and so on), which was (and is) due to the experience of the designers. Only recently, Python has been proposed as an efficient paradigm in performing rapid design space exploration [11]. Indeed, after a careful investigation of the various high level languages that could be used in a flexible circuit HDL generation procedure (like C, C++, Java and so on), we selected the Python, due to some key elements of it. First, there are no restrictions as to the data that can be stored in a variable. In Python every variable has unlimited range; the variables do not have a fixed size, as in the other high level languages, like C or Java. We had faced this problem some years ago, as we were developing another CAD tool for Residue Number System, when we were asked to use our C based tool to generate HDL for large bitwidths; some variables overflowed and the results were erroneous. On the other hand, using Python ensures us that no overflow will happen to a variable of this language. Second, Python code is easy to be maintained, due to the clean structure. Third, Python is portable and can be executed (or better interpreted) in any operating system; in fact we develop the code in the Linux operating system, and the same code is deployed automatically on our FreeBSD web server. Fourth, there are many modules and libraries that can provide assistance to various tasks. Fifth, it is easy to use performance optimized C procedures for critical functions, and sixth, it is easy to learn, so people without any knowledge of Python can join our team for some period and develop specific parts of our tool. We believe in the future more EDA people will turn to Python to automate their tasks, in order to increase their productivity. The only drawback is the fact that it is an interpreted language, and for this reason, sometimes, it exhibits higher execution times, than similar tools in C.

The front end is a web based form, in which the user can give the requested design parameters of bitwidth of each input operand, the option to pipeline or not the circuit, and the number of testbenches to be created. This front end passes all the design parameters to the tool's back end.

The tool's back end provides the core functions, and for this reason we will focus only on it. This back end consists of three modules: ($i$) the Multiplier design module, which analyzes the user inputs and creates the specific design description in a special netlist format called $\alpha$-HDL, ($ii$) the HDL Generator module, which takes as input this netlist format and creates signals, networks, assignments, and connections, resulting in the output description in VHDL, and ($iii$) the VHDL Test bench creator, which takes as input the constructed data structures of the previous module, and generates a full VHDL test bench, with handles for automatic design validation.

---

[1]The tool is available at http://arch.icte.uowm.gr/hdl/multiplier_ahdl.php.

The generated circuits have the top-level diagram of Figure 1. The first building block is the partial sum generation, which accepts the two vectors of dimensions $n$ and $k$ bits and uses one bit AND gates of two inputs. The second building block is the carry save adder tree, which sums all the partial sums, using full adders and half adders, strategically placed in order to minimize the latency and the total number of components. The final building block is the ripple carry adder, which outputs the final multiplication result of $n + k$ bits.

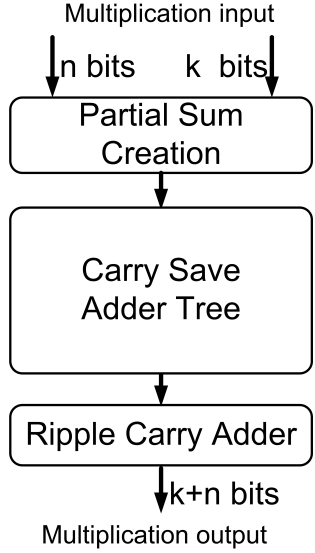

Figure 1. The top-level diagram of our generated circuits, where the three main building blocks (partial sum generation, carry save addition, ripple carry addition) are illustrated.

### A. Multiplier design module

The first module is the Multiplier design module, which creates a netlist in an internal format developed at our laboratory, which we call it $\alpha$-HDL format [6], and operates in three stages: $(a)$ partial sums creation, $(b)$ carry save addition, $(c)$ ripple carry addition. This module can be used to create multiplication units of unsigned vectors for arbitrary bit lengths. Also, due to the fact that we use the Python language, there are no restrictions as to the bitwidths of the vectors to be multiplied. For example the Xilinx Core generator can only create multiplication units up to $64 \times 64$ bits. Our tool has been used to create multiplication units with vectors up to $512 \times 512$ bits. Such large vectors are usually found in cryptographic applications [13].

**First Stage:** The first stage computes the network of AND gates, which create the bits that should be summed for every column. This stage accepts as input the bitwidth of the two input vectors. Given the two vectors, the module uses a number of one bit AND gates to compute the partial sums for each column, according to the standard multiplication rules. For example, if the bitwidth is two for each vector, then the two inputs are $X_0 X_1$ (the left bit is the LSB bit), and $Y_0 Y_1$. The multiplication of these two vectors will result in the following operations: A bit will be created in the less significant column (bit position 0) resulting from the $X_0 \otimes Y_0$, with $\otimes$ denoting the boolean AND operation, while in the next column (bit position 1) two bits will be created resulting from the operations $X_0 \otimes Y_1$ and $X_1 \otimes Y_0$. At bit position 2, one bit will be created resulting from the $X_1 \otimes Y_1$. The multiplication of 2-bit vectors will result in a 4-bit vector, which is taken into consideration in the next stage. The outcomes of the first stage are two: (a) the $\alpha$-HDL structure that defines the AND structures that create the bits, and (b) a two dimensional array that specifies for every column the bits that should be taken into consideration.

**Second Stage:** The second stage, accepts as input the array created in the previous stage and performs an optimized addition, using carry save adders. We have named this stage with the term reduction stage. This stage consists of many iterations. In every iteration $i$ the reduction stage, scans all columns $j$ starting from the least significant column, locates the columns that have more than one bit and places full adders (FA) or half adders (HA). The placement of adders is done in the best efficient way, in order to minimize the total number of FAs or HAs, as follows: Until the total number of bits to be added are over 2, full adders are placed in the netlist, with their output carry registered for future processing at the next iteration $(i + 1)$, at the next column $(j + 1)$, and output sum registered for future processing at the next iteration $(i + 1)$, same column $(j)$. If the number of bits to be added is 2, then the tool examines whether to add a HA, or to delay the insertion of the HA in favor of a better utilization in a future iteration. The tool will not add an HA when a carry has been registered at the next iteration $(i + 1)$, for this column $(i)$, because a FA can be used in the next iteration to add all three bits. Also, the tool will not add an HA when two bits have been registered at the next iteration $(i + 1)$, of the previous column $(j - 1)$, because in the iteration $(i + 1)$ a carry bit will be created and be registered at iteration $(i + 2)$, column $(j)$. Thus, in iteration $(i + 2)$, a FA can be used to add all three bits. When the total number of bits in a column is less than 2, the reduction stage completes. As it can be observed the placement of a FA or HA is done in a strategical way, that guarantees that the minimum amount of hardware resources will be used.

**Third Stage:** The third stage of the multiplier design module, is the final addition using a ripple carry adder. This stage, which is also optimized, places the best number and types of adders. This is done by checking the total number of bits to be added in every column (0 or 1 or 2), and then decides whether to direct connect the column to the output (when the bit is 0 or 1 and no carry has been generated in the previous column), to place a HA (when this column carries two bits and no carry bit has been generated in the previous column, or this column carries one bit and a carry bit has been generated in the previous column), or to place a FA (when 3 bits have to be added).

The multiplier module also accepts as input the option to pipeline the design or not. The pipelined design uses D flip flops (DFF) to delay input and output bit columns, and increases the throughput of the design, with the cost of increased hardware (Section IV). The tool carefully adds delay units both to inputs and output columns, for a uniform delay to every bit. Currently, the user cannot select the pipeline stages, and a DFF is added in every iteration, maximizing the operating frequency and resulting in a optimum circuit for performance.

The critical path using pipeline is equal to two gates delay (full adder module).

The output of the multiplier design is a structure in the $\alpha$-HDL netlist format, which is then processed by our HDL Generator Module and creates fully synthesizable VHDL description (Section III-B), following by the test bench generator (Section III-C).

*B. HDL Generator Module*

The netlist created in the previous stage is given as input to the HDL Generator Module. This is a general purpose VHDL generator library that can be easily connected to many different generators. This module accepts as input a special and compact netlist format, which we name it abstracted HDL $\alpha$-HDL [6]. This netlist format, as well as the HDL Generator Module, do not belong to the scope of this paper, and thus we will not describe them further.

*C. HDL Test bench Generator*

This module, is of out most importance, because it creates multiple vectors of testbenches, which can be used to test the correctness of the design in an HDL simulator. As it is evident, the multiplier hardware design module is a very complicated process, which should be tested thoroughly. Our tool accepts as input the number of input cases to create, and generates the test bench file in VHDL file. To do this, first it creates an empty entity declaration, then it instantiates the top level component and creates signals for every input and output port. Furthermore, it creates a clock process and a function that is used to convert bits to integer. The next step is to create the requested number of input test cases.

For the number of input test cases, the following loop is performed: for every operand a random number is created and converted to a binary, and extended to the full bitwidth of the operand. For example, if the random number 3 is selected with bitwidth 8, then this vector will be 00000011.

The tool multiplies both vectors and precomputes the final result. After the value assignments, it inserts a wait clause for the delay, which was computed in the previous stage, and then constructs an 'assert' statement to check the output. For example, if one 8 bit vector has the random binary value 00110101b (53), and another has the random value 00010111b (23), resulting in 1219, then the generated VHDL test bench file will carry the lines shown in Figure 2. As it can been seen in this figure, if the output is not the correct (1219) after the wait-time that has been computed before and corresponds to the circuit's latency, a message will be printed with severity error that will specify the input vectors, the expected result and the computed result. In every other case, a success message will be printed. In our web tool, the user can select the number of test vectors that will be automatically created.

All the test bench vectors are created randomly and automatically, according to the requested number of tests by the user. Also, all the checks are done automatically, which means that the designer can load the test bench file into his HDL Synthesis and Simulator tool, and can execute it without any other intervention.

IV. EXPERIMENTAL RESULTS

In order to evaluate the efficiency of our web tool, we generated a large number of VHDL descriptions for different

```
-- input vector: 53
signal0<="00110101" AFTER  17 ns ;
-- input vector: 23
signal1<="00010111" AFTER  17 ns ;
wait for waittime * 1 ns ;
-- output: 1219
assert (vec2int(signal4) = 1219  )
report "TESTBENCH Output:
"&integer'image(vec2int(signal4))&"
Expected:"&integer'image(1219)
severity error ;
assert (vec2int(signal4) /= 1219  )
report "TESTBENCH OK" severity note ;
```

Figure 2. Sample automated test bench created by our tool, for multiplier verification.

design parameters.

Even though the circuits were designed with a 'correct-by-construction' method, we had to verify the correctness and the structural integrity of the generated codes, with an automatic way. To do this, we followed an iterative procedure of three steps: generate, verify, modify. For the generation step, we used a verification script that created multiplication circuits of input bitwidth 1 to 32 by input bitwidth 1 to 32 (1024 circuits), and for each circuit up to 100 unique and random test vectors were created by our test bench generator tool (Subsection III-C). Thus, there were circuits that were verified completely, usually for small input bitwidths, and circuits that were tested in a significant percentage but not completely, due to lack of time. For the testing step, the VHDL command line compiler and simulator ghdl[2] was used to simulate the generated circuit. The delivered circuits are bug free, which can be verified by downloading the random testbench file from our site. Except the testbenches, the quality of the generated descriptions is verified partly by our HDL tool that accepts the $\alpha$HDL netlist format and creates the VHDL codes. The tool, performs many connectivity tests on every component, and in case something is not connected, then a specific warning is displayed.

We have to note that even though multiplication is a common operation at arithmetic circuits, there is no tool available from the web that can generate HDL codes of multipliers for different operand bitwidths. As discussed, the only online and related tool is [2], which bears the shortcomings discussed in the Related Work section (II). Furthermore, both Xilinx and Altera provide a tool to create a parametrized multiplier, though the outcome is not a VHDL file but a binary encrypted implementation netlist, which can be used only in a project targeting a specific FPGA board family. In contrast with these two vendors, our tool creates generic VHDL code that is vendor neutral and can be freely synthesized either in FPGA or in ASIC.

Another remark is, that even though there are few offline tools to create multiplication cores, all these cores do not use carry save adders to compute the product, but they use special structures that make use of fast DSP blocks found on modern FPGA boards. Thus, on the one hand we cannot provide estimations with other CSA multipliers, and on the other hand, comparing our CSA multiplier scheme, with other schemes of multiplication can be used only to extract some general conclusions and not to determine the efficiency of our circuits. Nevertheless, we have performed such comparisons and we provide the results in this section.

---

[2]https://gna.org/projects/ghdl/

Some of our design metrics are summarized on Table I. Specifically, the *testcase* column defines the input bitwidths of the two vectors. Accordingly, the signals, adders and DFF, show the number of internal signals used, the total quantity of the adders, and the D flip flops. This table illustrates the importance of a tool to generate syntactically and operationally correct hardware descriptions of circuits, especially for non-trivial bitwidths and inputs.

Table I. AUTOMATIC GENERATED MULTIPLIER CODES

| #testcase | signals | delay | adders | DFF |
|---|---|---|---|---|
| 8x8 | 203 | 17 | 62 | 227 |
| 16x16 | 1858 | 33 | 254 | 1061 |
| 32x32 | 7721 | 65 | 1023 | 4588 |
| 128x128 | 61001 | 115 | 16105 | 24104 |

Concerning the runtime of the tool, we provide the Table II, which shows the execution time of each module and the maximum runtime memory footprint, for different input vectors, running on Intel Core2 Duo E7600 3.06 GHz with the FreeBSD operating system. Our EDA tool has very low memory requirements, even for complicated design parameters, which is important for a web application, where many users are expected to use it simultaneously. Also, the execution time is reasonably low for normal design parameters, but it requires a fair amount of minutes, in case very large input vectors are requested. It should be noted that the test bench module always creates the test bench in less than 2 seconds, while the module that creates the schematic has an execution time linear to the product of number of components times signals. Comparing with the Xilinx *Core Generator* the runtime is similar, but our tool, being a console application, has a much lower memory footprint (for example where our tool required 40 MB the Xilinx core generator required 370MB). Of course the Xilinx tool provides more generic functionality and for this reason it puts more memory pressure on the system. On the other hand, if the designer would like only the generation of one or more multiplication units using the Xilinx tool, he cannot opt out of this extra functionally and thus he is required to provide this great amount of memory. Our Tool on the other hand, requires minimum amount of memory on the designers client workstations, because it only needs an Internet browser.

Table II. PERFORMANCE METRICS OF THE WEB TOOL

| #testcase | #Mul(s) | HDL(s) | Mem(MB) |
|---|---|---|---|
| 8x8 | 1 | 1 | 35 |
| 16x16 | 1 | 1 | 38 |
| 32x32 | 1 | 2 | 40 |
| 32x64 | 2 | 4 | 45 |
| 64x64 | 200 | 25 | 50 |
| 64x128 | 1501 | 305 | 52 |
| 128x128 | 6205 | 1010 | 66 |
| 128x256 | 12003 | 1520 | 95 |
| 256x256 | 16040 | 2300 | 140 |

Additionally, we synthesized the generated VHDL codes with Leonardo Spectrum, Xilinx Vivado 2013.2, Altera Quartus II 12.0. The synthesis results (Figure III) from Xilinx Vivado (Virtex6, speed grade -2) show that for small input bitwidths (vectors of 8x8), the occupied slices for the pipeline version (denoted with the letter 'p') are low, and in case that we did not pipeline them, the usage is extremely small (5 slices only). As the input bit widths are increased the occupied slices follow a similar trend, but the maximum operational frequency remains stable and over 500 Mhz. On the same Table, we have included the total power consumption for every test case, while on Table IV we give the detailed breakdown of it. The power estimations were performed using the Xilinx Xpower Analyzer, for the part xc6vlx760, package ff1760 Virtex6 FPGA implementation, with settings of Ambient Temperature 50 °C, Medium Profile Heat Sink, 12 Board Layers, Supply Voltage 2.5V, and frequency target the maximum operational frequency of each circuit. Due to lack of space we do not provide the metrics of Altera Quartus or Leonardo Spectrum, which are similar.

Table III. SYNTHESIS RESULTS OF OUR MULTIPLICATORS

| #testcase | #Freq(Mhz) | Slices | Power(W) |
|---|---|---|---|
| 8x8 (p) | 589.970 | 116 | 4.803 |
| 8x8 | 330 | 5 | 4.450 |
| 16x16 (p) | 589.970 | 444 | 5.218 |
| 16x16 | 90 | 124 | 4.461 |
| 32x32 (p) | 538.213 | 1125 | 5.501 |
| 32x32 | 62 | 590 | 4.473 |
| 63x13 (p) | 538 | 1350 | 5.252 |
| 63x13 | 53 | 566 | 4.474 |

Table IV. DETAILED POWER ANALYSIS OF OUR GENERATED CIRCUITS

| #testcase | Clock(W) | Logic (W) | Signals(W) | IOs(W) | Leakage(W) |
|---|---|---|---|---|---|
| 8x8 (p) | 0.080 | 0.005 | 0.008 | 0.245 | 4.465 |
| 8x8 | 0 | 0.001 | 0.001 | 0.001 | 4.447 |
| 16x16 (p) | 0.147 | 0.025 | 0.030 | 0.531 | 4.486 |
| 16x16 | 0 | 0.001 | 0.008 | 0.005 | 4.447 |
| 32x32 (p) | 0.248 | 0.080 | 0.091 | 0.581 | 4.500 |
| 32x32 | 0 | 0.002 | 0.017 | 0.006 | 4.448 |
| 63x13 (p) | 0.255 | 0.060 | 0.074 | 0.376 | 4.488 |
| 63x13 | 0 | 0.002 | 0.017 | 0.007 | 4.448 |

As we mentioned in the beginning, a similar offline tool like ours is Flopoco [7], which it's main target is floating point arithmetic, but it can also generate autonomous integer arithmetic circuits for operands with different input bitwidths used in the floating point arithmetic, like integer multiplication. We used this tool to create a number of indicative circuits (Table V) for integer multiplication units, and compare them with our own generated circuits (Table III). The Flopoco tool creates the multiplication units using a number of compressors and structures specific to FPGA slices. For example it utilizes the DSP block $16 \times 16$, found on modern Xilinx FPGAs. Thus, the area requirements are very low. On the other hand, our circuits may consume many more slices, but perform faster for every input requirements, either the pipelined or the non-pipelined version; with other words, our generated circuits trade area for speed. Especially for lower input requirements (e.g. $8 \times 8$) our circuits operate at four times higher frequencies. Another difference of our tool and Flopoco, is the generation of the block schematic from our tool; the Flopoco does not generate any schematic. Finally, all of our circuits and schematics are generated by the web interface, without any requirement for local installation, and all generated circuits are accompanied by a random HDL test bench file to verify their correctness (the Flopoco does not provide a test bench file).

Finally, we used the proprietary tool of Xilinx named *Core Generator*, to generate multiplication cores optimized for speed. Table VI illustrates the results of some of the circuits. It is evident that the generated cores are very efficient concerning the occupied slices, but again the maximum frequency is lower than our generated cores. On the other hand our generated cores require many more slices, but given the fact that in a contemporary FPGA chip like Virtex 6 (xc6vlx760) there are over 110.000 slices, our area requirements of 1000 slices can be considered negligible (less than 1%). Thus, our circuits are faster than Xilinx, but consume more slices. On the Table VI the Xilinx tool was not able to report power estimations, because the generated cores are encrypted and can only be embedded on a bigger circuit; they cannot be modeled stand alone, as was the case with our's or Flopoco's vendor neutral HDL.

Table V. SYNTHESIS RESULTS OF FLOPOCO

| #testcase | #Freq(Mhz) | Slices | Power(W) |
|---|---|---|---|
| 8x8 (p) | 87 | 1 | 4.540 |
| 8x8 | 87 | 1 | 4.451 |
| 16x16 (p) | 200 | 1 | 4.595 |
| 16x16 | 83 | 1 | 4.455 |
| 32x32 (p) | 425 | 87 | 5.283 |
| 32x32 | 55 | 28 | 4.479 |
| 63x13 (p) | 421 | 24 | 4.968 |
| 63x13 | 51 | 17 | 4.459 |

Table VI. SYNTHESIS RESULTS OF XILINX GENERATED MULTIPLIER

| #testcase | #Freq(Mhz) | Slices |
|---|---|---|
| 8x8 (p) | 417 | 1 |
| 16x16 (p) | 450 | 1 |
| 32x32 (p) | 450 | 4 |
| 63x13 (p) | 245 | 16 |

## V. CONCLUSIONS

Design automation and fast circuit verification are the foundations for increasing productivity and achieving the fast time-to-market constraints for heterogenous computing. Multiplication is a fundamental operation in every digital circuit, and especially at DSP. Here, we present our contribution to the EDA domain, by providing a tool accessible from the web, to generate syntactically correct multiplication units in the VHDL language. The tool outputs the synthesizable VHDL description, a custom and automated test bench, block schematic, and other metrics. It is available for every anonymous user over the Internet. Synthesis results indicate the high performance on the Xilinx Virtex 6 FPGA, with operational frequencies up to 589Mhz. The generated code can be synthesized in FPGA or in ASIC projects, and it is not vendor specific.